# Orientation Dependence of Ferroelectric Properties of Pulsed-Laser-Ablated $Bi_{4-x}Nd_xTi_3O_{12}$ Films

A. Garg[a)] and Z.H. Barber

Dept. of Materials Science and Metallurgy, Pembroke Street, University of Cambridge, Cambridge CB2 3QZ, UK

M. Dawber and J.F. Scott

Dept. of Earth Sciences, Downing Street, University of Cambridge, Cambridge CB2 3EQ, UK

A. Snedden and P. Lightfoot

School of Chemistry, University of St. Andrews, St. Andrews, KY16 9ST, UK

**Abstract:**

Epitaxial (001)-, (118)-, and (104)-oriented Nd-doped $Bi_4Ti_3O_{12}$ films have been grown by pulsed-laser deposition from a $Bi_{4-x}Nd_xTi_3O_{12}$ (x=0.85) target on $SrRuO_3$ coated single-crystal (100)-, (110)-, and (111)-oriented $SrTiO_3$ substrates, respectively. X-ray diffraction illustrated a unique epitaxial relationship between film and substrate for all orientations. We observed a strong dependence of ferroelectric properties on the film orientation, with no ferroelectric activity in an (001)-oriented film; a remanent polarization, $2P_r$, of 12 µC/cm$^2$ and coercive field, $E_c$, of 120 kV/cm in a (118)-oriented film; and $2P_r$ = 40 µC/cm$^2$, $E_c$ = 50 kV/cm in a (104)-oriented film. The lack of ferroelectric activity along the c-axis is consistent with the orthorhombic nature of the crystal structure of the bulk material, as determined by powder neutron diffraction.
85.50$^2$.-n, 77.84.-s, 77.55.+f

[a] Present address: Dept. of Materials and Metallurgical Engg., Indian Institute of Technology, Kanpur - 208016, India; electronic mail: ashishg@iitk.ac.in

SrBi$_2$Ta$_2$O$_9$ (SBT) films were the first of the Bi-layered oxides to show promise for Ferroelectric Random Access Memory applications, due to their excellent fatigue resistance on Pt electrodes[1]. However, growth of SBT films requires high temperatures (> 750°C, and typically 850°C or higher for epitaxial films[2,3,4]), and films show low remanent polarization, P$_r$ (typically 2P$_r$ ≈ 10 µC/cm$^2$ on Pt). The emergence of La-doped bismuth titanate (Bi$_{4-x}$La$_x$Ti$_3$O$_{12}$ or BLT)[5] as a possible substitute fuelled considerable research to find other variants and further enhance film properties. BLT can be deposited at lower temperatures (≤ 750°C) on Pt electrodes and shows a higher 2P$_r$ (approx. 24 µC/cm$^2$) than SBT, as well as excellent fatigue resistance.

Bi$_4$Ti$_3$O$_{12}$ is an Aurivillius phase Bi-layered oxide, and can be denoted by the formula (Bi$_2$O$_2$)$^{2+}$ (Bi$_2$Ti$_3$O$_{10}$)$^{2-}$, in which perovskite units of Ti-O octahedra are sandwiched between Bi$_2$O$_2$ layers. Bulk undoped Bi$_4$Ti$_3$O$_{12}$ shows a very high 2P$_r$ (about 100 µC/cm$^2$)[6], but thin films have much lower values of switching polarization and suffer from fatigue upon bipolar switching[7]. It was proposed that doping with La led to improved oxygen ion stability in the lattice and hence improved fatigue resistance because some of the Bi ions in the pseudoperovskite layers containing Ti-O octahedra were substituted by La ions. Substitution of the non-spherical Bi$^{3+}$ cation with La$^{3+}$, however, reduces the structural distortion of the perovskite block, thereby reducing P$_r$. Substitution by lanthanide ions having smaller ionic radii than Bi, such as Nd or Sm (comparative ionic radii for eight-coordination: Bi$^{3+}$ 0.117nm, La$^{3+}$ 0.116 nm, Nd$^{3+}$ 0.111, Sm$^{3+}$ 0.108 nm) should maintain a more significant structural distortion and improved ferroelectric properties, in particular P$_r$.

Recently, Chon *et al.*[8] have reported very high values of remanent polarization in sol-gel derived lanthanide-doped bismuth titanate thin films, with 2P$_r$ values as high as ~100 µC/cm$^2$ in highly *c*-axis oriented Nd-doped (BNdT) films. The authors attributed this, rather unexpected, result to the development of the polarization vector along the *c*-axis of the BNdT unit-cell: in undoped Bi$_4$Ti$_3$O$_{12}$ the polarization is almost exclusively along the *a*-axis. Based on this explanation, we would expect BNdT films of the same composition with orientations further away from the *c*-axis (or closer to the *a*-axis) to exhibit lower P$_r$. In order to investigate this issue we have deposited epitaxial BNdT films and measured the dependence of P$_r$ upon film orientation.

In this letter, we report on the properties of epitaxial BNdT films of various crystallographic orientations (*c*-axis and non-*c*-axis oriented), grown by Pulsed Laser Deposition (PLD) from a target of composition Bi$_{3.15}$Nd$_{0.85}$Ti$_3$O$_{12}$ on single crystal (100)-, (110)- and (111)-oriented SrTiO$_3$ substrates. BNdT films, 300 nm thick, were deposited in 200 mtorr oxygen, at a substrate temperature of 750°C, and were subsequently cooled to room temperature (in approx. 45 minutes) at an oxygen pressure of ½ atm. The laser fluence during the deposition was approximately 2.8 J/cm$^2$ at a repetition rate of 5 Hz. The target was fabricated by cold pressing and sintering (1100°C for 1 hour in air) a powder compact prepared by a conventional solid-state reaction route. Prior to BNdT film deposition, conducting 50 nm thick SrRuO$_3$ films were deposited on the SrTiO$_3$ substrates to act as a lower electrode: these films were deposited at 650°C in 110 mtorr oxygen (laser fluence = 2.5 J/cm$^2$ approx.; repetition rate = 4 Hz). Pseudo-cubic SrRuO$_3$ has a very low lattice mismatch with SrTiO$_3$ and hence maintains the substrate orientation.

The structure of the films was studied by X-ray diffraction (XRD). For ferroelectric characterization Pt top contacts (area = 8.1 x 10$^{-5}$ cm$^2$) were deposited by sputter deposition through a shadow mask. Ferroelectric properties were measured using a Radiant Precision Pro. ferroelectric tester. Crystallographic studies were also made on bulk powder specimens by neutron diffraction.

A BNdT film on SrTiO$_3$ (100) was highly *c*-axis- or (001)-oriented, as shown by very strong 00*l* peaks on an XRD 2θ-scan (Figure 1 (a)). A (117) pole figure (Figure 2 (a)) shows four-fold symmetry with four sharp peaks at ψ ≈ 50° (angular spread < 5°), illustrating a very highly *c*-axis-oriented film with excellent in-plane orientation. The orientation relationship between the film and substrate is (001)$_{film}$ || (100)$_{substrate}$ and [100]$_{film}$ || [110]$_{substrate}$.

A BNdT film deposited on SrTiO$_3$ (110) was (118)-oriented, as shown by the XRD 2θ-scan in Figure 1 (b) and confirmed by a (117) pole figure scan, showing two sets of three peaks at ψ ≈ 5°, 65° and 85° corresponding to 117, 1$\bar{1}$7/$\bar{1}$17, and $\bar{1}\bar{1}$7 reflections (Figure 2 (b)). These two sets of peaks, related by mirror symmetry, indicate the presence of double-twin in the film, as previously reported for BLT on a similar substrate[9]. This twinning is due to two possible *c*-axis growth directions of BNdT perpendicular to the (100) planes of SrTiO$_3$, which are at 45° to the (110) substrate plane.

A BNdT film deposited on SrTiO$_3$ (111) was (104)-oriented as shown by the 2θ-scan in Figure 1 (c). It should be noted that only the 014 peak is present, as 104 is prohibited due to systematic absences. A (117) pole figure of this film showed a three-fold symmetry with three sets of peaks at ψ ≈ 35° (117 and 1$\bar{1}$7 reflections) and 85° (11$\bar{7}$ and 1$\bar{1}\bar{7}$ reflections), confirming good quality in-plane orientation. This pole figure can be interpreted in terms of the formation of three in-plane orientations due to *c*-axis BNdT growth along three directions (perpendicular to the SrTiO$_3$ (100) planes), separated by azimuthal angles of 120° leading to a triple-twin situation, also reported for BLT films on SrTiO$_3$ (111) substrates[9].

These results show that BNdT grows with a unique epitaxial relationship to the underlying $SrRuO_3$/$SrTiO_3$ substrate, with the *c*-plane of BNdT parallel to the (100) plane of $SrTiO_3$, as previously reported for BLT film growth[9] and maintaining a similar orientation relationship on all substrate orientations.

Figure 3 shows the ferroelectric hysteresis loops for the three film orientations described above. A *c*-axis-oriented BNdT film (on (100) $SrTiO_3$) did not show any ferroelectric activity, as demonstrated by a linear dielectric response and the absence of any hysteresis loop in Figure 3 (a). A (118)-oriented film (on (110) $SrTiO_3$) showed a $2P_r$ of ~12 $\mu C/cm^2$ and coercive field, $E_c$ of 120 kV/cm at an applied voltage of 10 V and a frequency of 200 Hz (Figure 3 (b)), whilst a (104)-oriented film (on (111) $SrTiO_3$) showed a $2P_r$ value of approximately 40 $\mu C/cm^2$ and an $E_c$ of 50 kV/cm using the same parameters. Results for the (104)-oriented film are in agreement with the results obtained by Kojima *et al.*[10] for BNdT films grown by metalorganic chemical vapor deposition.

Although we have shown that doping with Nd leads to an increase in $P_r$ as compared to La-doped films, due to increased lattice distortion, we do not see the dramatic increase in the remanent polarization along the *c*-axis, as observed by Chon *et al.*[8]. On the contrary, our results suggest that $P_r$ is zero in *c*-axis oriented films, and the polarization increases on moving away from the BNdT *c*-axis.

In support of our thin-film results we have carried out a careful Rietveld analysis of the crystal structure of our $Bi_{3.15}Nd_{0.85}Ti_3O_{12}$ sample in bulk polycrystalline form using powder neutron diffraction[11]. This analysis clearly shows that $Bi_{3.15}Nd_{0.85}Ti_3O_{12}$ adopts an orthorhombic structure, space group B2cb, rather than the higher symmetry tetragonal possibility recently suggested by Chon *et al.*[12]. The orthorhombicity is confirmed by the presence of key superlattice reflections such as 014, 125 and 034; no significant peak splitting is observed, due to the similarity of the *a/b* lattice parameters. Moreover, there is no evidence for reduction of crystal symmetry to monoclinic (although we cannot rule out a very small distortion), as suggested for $Bi_4Ti_3O_{12}$ itself. These subtle features are key, since both tetragonal and monoclinic groups naturally allow a c-axis component to $P_r$, whereas the orthorhombic B2cb model necessitates only an *a*-axis polarization. Although there is evidence that the bulk and thin-film crystal structures of ferroelectric materials may differ due to epitaxial constraints (e.g. $BiFeO_3$ thin films[13]), in this case we have direct evidence for the preservation of orthorhombic symmetry in the thin-film – the presence of the 014 peak in Fig. 1(c) negates tetragonal symmetry.

Although the results of Chon *et al.* remain unexplained, the situation in strontium bismuth niobate (SBN) described by Watanabe *et al.*[14] may be analogous; in that case they found that small deviations from stoichiometry, linked to preferred orientation, led to changes in Pr along the c-axis by as much as 70%.

Another interesting observation from these results is that the coercive voltage of the (118)-oriented film is much larger than (104)-oriented film which was not observed in the BLT thin films[9]. Reasons for this are not clear at the moment but one possibility may be that Nd doping can lead to different magnitudes of distortions along different crystallographic directions due to the size difference between Nd and La, making some directions more easily switched than others.

In summary, we have found that (001)-oriented BNdT films deposited by PLD do not show ferroelectric activity. A (118)-oriented film showed a $2P_r$ value of 12 $\mu C/cm^2$ at a coercive field of 120 kV/cm, whereas a (104)-oriented film had a much higher $2P_r$ value of ~40 $\mu C/cm^2$ and lower $E_c$ of ~50 kV/cm. These results are consistent with the proposed model of the crystal structure for $Bi_{3.15}Nd_{0.85}Ti_3O_{12}$ derived from neutron diffraction results.

AG would like to thank Downing College, Cambridge for financial support during this work and Dr. A.Q. Jiang for his help with ferroelectric measurements.

**List of Figures**

**Figure 1**  2θ scans of BNdT films on SrTiO$_3$ substrates of various orientations: (a) (100)- (b) (110)- and (c) (111)-oriented. ● - SrTiO$_3$ K$_{\alpha 1}$, ◊-SrRuO$_3$ K$_{\alpha 1}$, *-unidentified.

**Figure 2**  (117) Pole figures of (a) (001)-oriented, (b) (118)-oriented and (c) (104)-oriented BNdT Film. Pole figures are plotted on Wulff Scale and each division corresponds to 10° along the radii (ψ) and 30° along the circumference (φ).

**Figure 3**  P-E Loops for (a) (001)-oriented, (b) (118)-oriented and (c) (104)-oriented BNdT film. The applied voltage was 10 V at a frequency of 200 Hz.

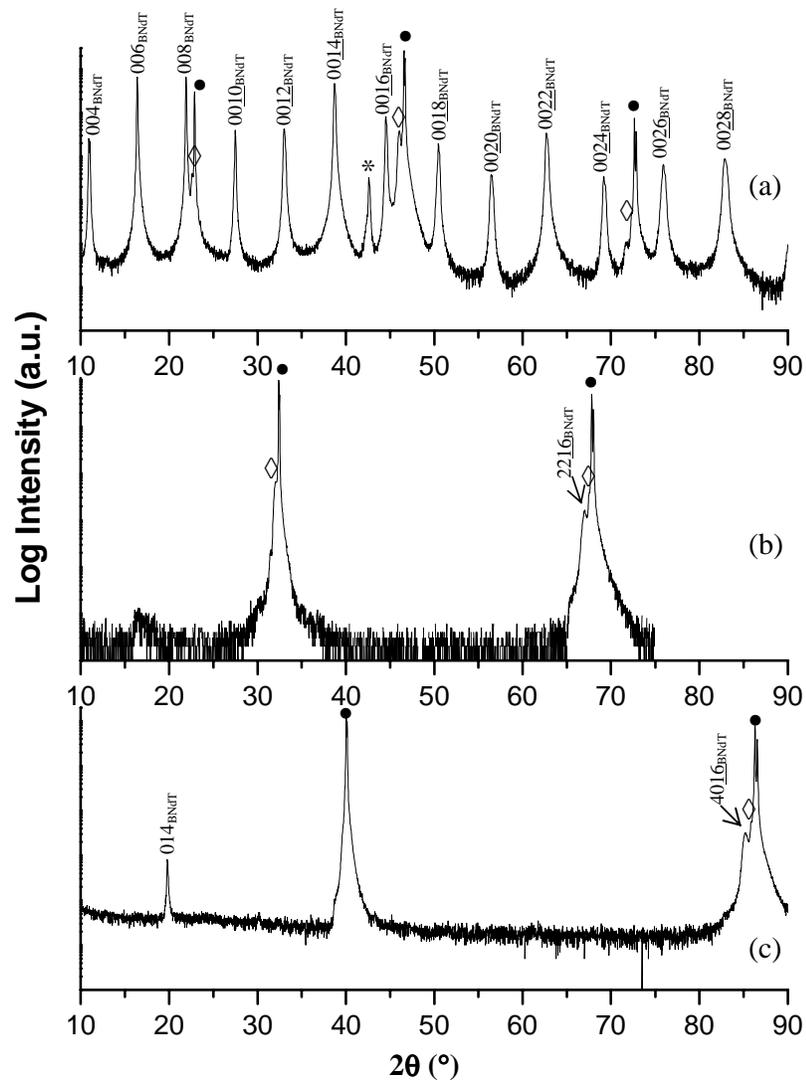

**FIGURE 1/3**

A. GARG *et al.*

Applied Physics Letters
12

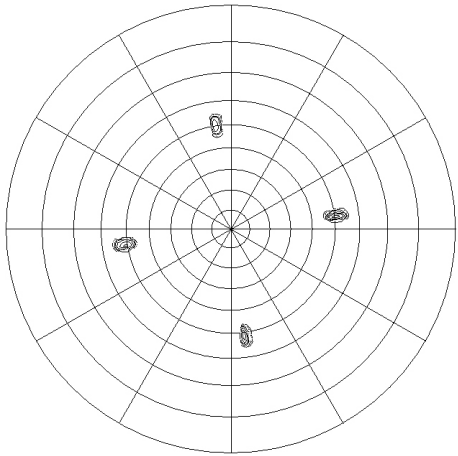 (a)

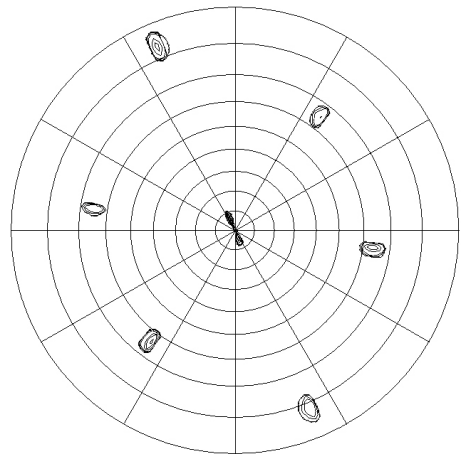 (b)

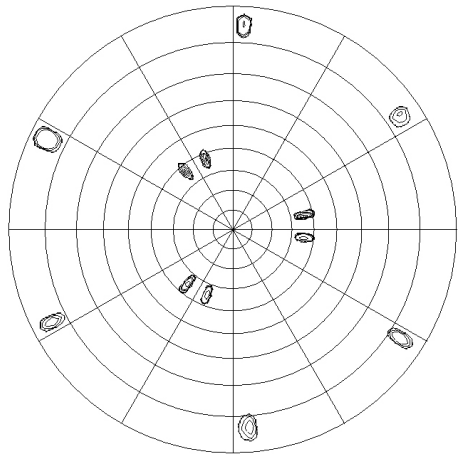 (c)

**FIGURE 2/3**

A. GARG *et al.*

*Applied Physics Letters*

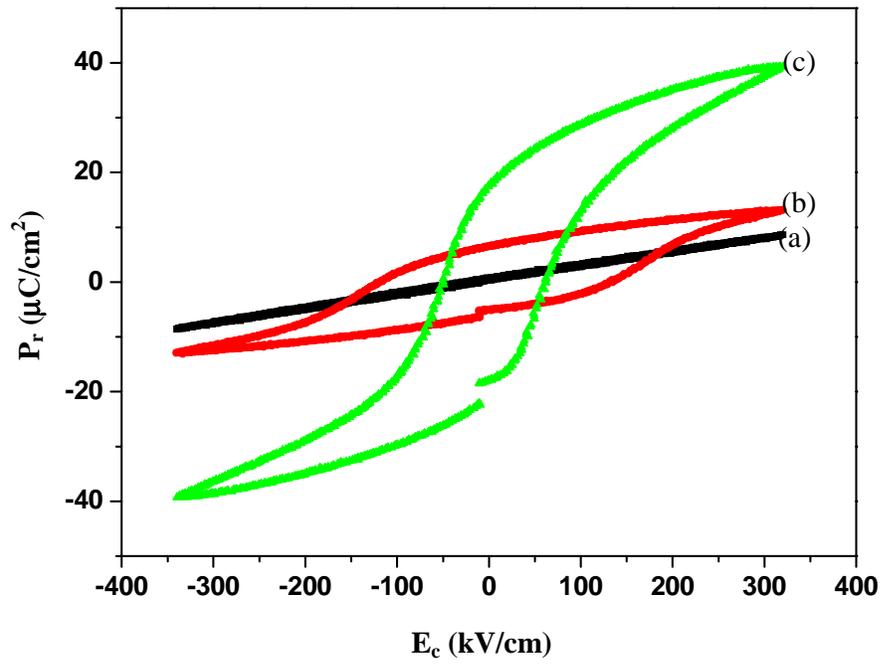

**FIGURE 3/3**

A. GARG *et al.*

*Applied Physics Letters*